\documentclass[a4paper,american,11pt]{article}
\usepackage{a4wide}
\usepackage[utf8]{inputenc}
\usepackage{babel}
\usepackage{authblk}
\usepackage{mathtools}
\RequirePackage{physics}
\usepackage{xparse}
\usepackage{graphicx}
\usepackage{caption}
\usepackage{subcaption}
\usepackage[title]{appendix}

\title{Charge transport battery with quantum feedback }
\author{Oscar Boh\'orquez}
\affil{Departamento de F\'{\i}sica y Geología,  Universidad de Pamplona}
\date{}

\begin{document}

\maketitle

\begin{abstract}
A battery is a work storage device, i.e. a device that stores energy in the form of work for later use by other devices. In this work, we study the realization of a quantum battery in a double quantum dot in series, charged by two electrodes at different chemical potentials and optimized by a Markovian quantum feedback protocol. Using the concept of ergotropy as a figure of merit, we first establish a simple expression for the maximum ergotropy in a two-level system, and then find the parameters under which a Markovian feedback can achieve this optimal ergotropy. We also study the influence of interaction with a phonon environment on the charging and discharging process of the battery.
\end{abstract}

\section{Introduction}

Quantum technologies promise to greatly enhance the performance of some current technologies; indeed, there have already been several notable successes in quantum metrology, cryptography, and quantum computing \cite{Acin2018, Joseph2023}.  In recent years, research on quantum batteries has attracted attention due to the need for energy storage \cite{Alicki2013, Binder2015, Campaioli2018, Ferraro2018, Barra2019, Gyhm2021}, both for the use of quantum devices and for more efficient storage methods with possible scalability to macroscopic devices. It is well known that quantum phenomena such as coherence and entanglement offer the possibility of improving the performance of technologies such as computing by several orders of magnitude, and the same can be expected for the energy storage capacity and charging time of quantum batteries. From the seminal paper by Alicki and Fannes showing that entanglement increases the amount of work extracted from a set of batteries \cite{Alicki2013}, to the recent experimental realization of a superconducting quantum battery \cite{Hu2022}, there has been an increasing research into how quantum phenomena can increase the efficiency of batteries. 
The process of charging and discharging a battery depends on three elements: the charger, which in the quantum case is typically an external field; the battery itself, which is the system that stores the energy; and the consumption center, which drains the battery using the energy as work.
Different models of quantum batteries have been studied, from theoretical analysis \cite{ Alicki2013, Binder2015, Andolina2018} to possible experimental realizations \cite{Ferraro2018, Hu2022, Liu2019}, both in single cells \cite{Andolina2018, Gherardini2020,  Monsel2020, tacchino2020charging} and in many-body models and their collective effects \cite{ Ferraro2018, Gyhm2021, Le2018, Rossini2020}. Also in closed systems \cite{Binder2015, Campaioli2017, Crescente2020} as well as in open systems \cite{Santos2021, Carrega2020, Cakmak2020, Farina2019}, the latter due to the inevitability of the interaction of the system with the environment and the resulting decoherence effects. Among the experimental proposals are: superconducting qubits \cite{Santos2019, Gemme2022}, cavity and circuit quantum electrodynamic architectures \cite{Ferraro2018, Carrasco2022} and spin batteries \cite{Le2018, Huangfu2021}. In this work we study a charge transport quantum battery in a double quantum dot coupled to two electronic reservoirs at different chemical potentials, under the scheme proposed by Pöltl et al. \cite{Poltl2011}, by showing how a Markovian feedback protocol can be used to optimize the battery charge; using the concept of ergotropy \cite{Allahverdyan2004} as a figure of merit. In the first part of this paper, a general expression for the ergotropy of a two-level system is obtained, and is discussed the role of quantum coherence; then we show how the application of a Markovian feedback \cite{Wiseman2009} can be used to reach the maximum ergotropy of the system. Finally, we analyze how the interaction with a phonon environment affects the charging and discharging of the system,  all within the framework of the Lindblad equation.

\section{Ergotropy of a two-level system}

\begin{figure}[ht]
\centering
\includegraphics[width=0.6\textwidth]{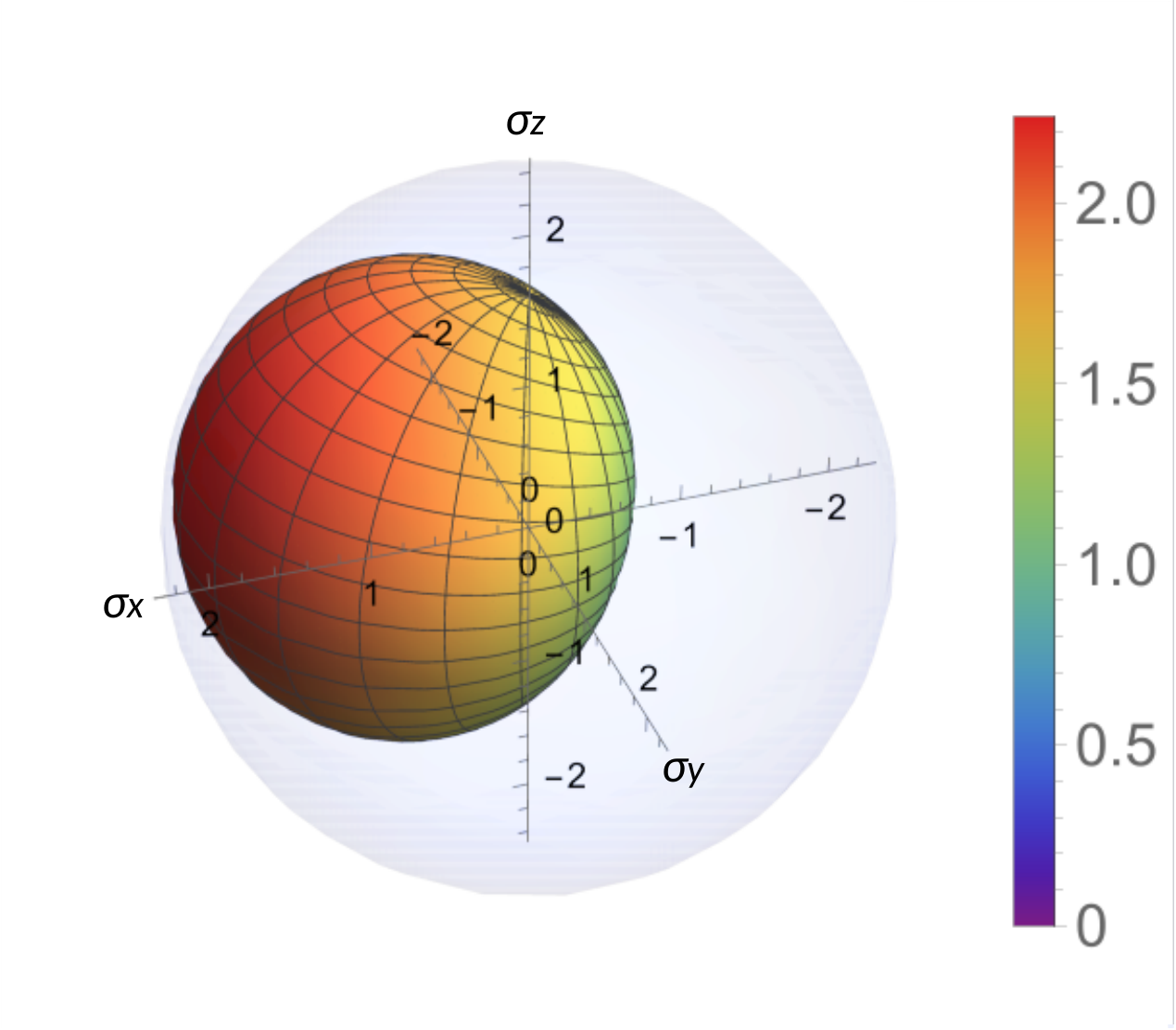}
\caption{Variation of the ergotropy in a two-level system in spherical coordinates with the parameters $\text{Tc}=\epsilon=1, r=1$.}\label{fig_1}
\end{figure}

Ergotropy is a concept introduced by Allahverdyan et al. \cite{Allahverdyan2004}, defined as the maximum amount of work that can be extracted from a finite quantum system by unitary operations. Given the initial state of a system $\rho_0$, work can be extracted by a potential $V(t)$ acting cyclically, i.e. it acts for a time $t$, otherwise it is zero. The final state is then found by solving the von Neumann equation
\begin{equation}
    \dot{\rho}(t)=-i[H(t),\rho(t)]\qquad \text{with } \qquad H(t)=H_0+V(t). 
\end{equation}   
where $H_0$ is the Hamiltonian of the system. The solution is
\begin{equation}
 \rho(t)=U(t)\rho_0U^\dag(t), \quad  U(t)=\hat{\tau}\exp\left(-i\int_0^t H(t')dt'\right),
\end{equation}
with $\hat{\tau}$ being the time-ordering operator. The ergotropy $\mathcal{W}$ is the difference in energy between the final state and the initial state, where the final state is one from which no work can be extracted; called a passive state $\rho(\tau)$:
\begin{equation}
\mathcal{W} =\operatorname{tr}\left(\rho_0 H_{0}\right)-\operatorname{tr}\left(\rho(\tau) H_{0}\right) \\
=\operatorname{tr}\left(\rho_0 H_{0}\right)-\operatorname{tr}\left(U(\tau) \rho_0 U^{\dagger}(\tau) H_{0}\right).
\end{equation}
Writing the initial state and the Hamiltonian in terms of their spectral decompositions, and keeping the order of the eigenvalues and their respective eigenvectors as shown below,
\begin{gather}
\rho_{0}=\sum_{j \geq 1} \rho_{j}|\rho_{j}\rangle\langle \rho_{j}|, \quad \rho_{1} \geq \rho_{2} \geq \cdots, \label{rho_spect} \\
H=\sum_{k \geq 1} \varepsilon_{k}| \varepsilon_{k}\rangle\langle\varepsilon_{k}|, \quad \varepsilon_{1} \leq \varepsilon_{2} \leq 
\cdots \label{h_spect},
\end{gather}
it can be shown that ergotropy can be expressed as
\begin{equation}\label{ergo}
\mathcal{W}=\sum_{j, k} \rho_{j} \varepsilon_{k}\left(\left|\left\langle \rho_{j} \mid \varepsilon_{k}\right\rangle\right|^{2}-\delta_{j k}\right),
\end{equation}
and the passive state reads $\rho(\tau)=\sum_{j} \rho_{j}\left|\varepsilon_{j}\right\rangle\left\langle\varepsilon_{j}\right|$.
In general, the state of a two-level system (qubit) can be represented by spherical coordinates on the Bloch sphere in the $\sigma_z$ basis, with $\ket{1}$ being the ground state and $\ket{0}$ being the excited state, according to the quantum information notation: 
\begin{equation}\label{gen_rho}
\begin{aligned}
\rho =\frac{1}{2}
 (1 + r \cos \theta ) \ket{1}\bra{1} + \frac{1}{2}r e^{-i \phi } \sin \theta  \ket{1}\bra{0} \\+
 \frac{1}{2}r e^{i \phi } \sin \theta \ket{0}\bra{1} + \frac{1}{2}(1-r \cos \theta )\ket{0}\bra{0}.    
 \end{aligned}
\end{equation}
where $r$ is is the norm of the Bloch vector, $\theta$ is the polar angle, and $\phi$ is the azimuthal angle. Thus, we can express the initial state $\rho_0$ in its spectral decomposition (\ref{rho_spect}), with the following eigenvalues and eigenvectors:
\begin{equation}
\begin{gathered}
\rho_1=\frac{1+r}{2}, \quad \rho_2=\frac{1-r}{2},  \\
|\rho_1\rangle=\cos (\theta/2)\ket{1} + e^{i \phi } \sin (\theta/2)\ket{0},\\
|\rho_2\rangle=
\sin (\theta/2)\ket{1}-e^{i \phi } 
\cos (\theta/2)\ket{0}. 
\end{gathered}
\end{equation}
In the same way, we can express the general Hamiltonian of a two-level system as:  
\begin{equation}\label{h_twolevel}
H = \frac{\epsilon}{2}\sigma_z+\text{Tc}\sigma_x,
\end{equation}
here $\epsilon$ is the energy difference (detuning) between the two levels and $\text{Tc}$ is the coupling between the levels, which can be the tunnel matrix element for a double well system, or due to an external magnetic field in a spin-1/2 system. This type of Hamiltonian is ubiquitous, and it is also found in superconducting qubits \cite{Leggett1987}. The spectral decomposition  of the Hamiltonian (\ref{h_spect}) is: 
\begin{equation}\label{h_eigen}
\begin{gathered}
   \varepsilon_1=-\frac{\Delta}{2}, \quad \varepsilon_2=\frac{\Delta}{2} \quad\text{with}\quad \Delta = \sqrt{4 \text{Tc}^2+\epsilon ^2},  \\
    |\varepsilon_1\rangle=\frac{1}{N_+}
    \Big( 2 \text{Tc} \ket{0} + (\epsilon - \Delta)\ket{1} \Big),
 \quad
|\varepsilon_2\rangle=\frac{1}{N_-}\Big( 2 \text{Tc}\ket{0} + (\epsilon + \Delta)\ket{1} \Big),\\
 \quad\text{with}\quad N_\pm=\sqrt{4 \text{Tc}^2+(\epsilon\mp\Delta)^2}.
 \end{gathered}
\end{equation}
Then, from equation \ref{ergo} it is straightforward to see that in spherical coordinates the ergotropy for a two-level system reads
\begin{equation}\label{gen_erg}
\mathcal{W}=\frac{1}{2} r\left(\Delta+\epsilon  \cos (\theta )+2\text{Tc} \sin (\theta ) \cos (\phi )\right),
\end{equation}
or in Cartesian coordinates of the Bloch sphere
\begin{equation}\label{gen_erg1}
\mathcal{W}=\frac{1}{2} \left(r\Delta+\epsilon\left\langle\sigma_{z}\right\rangle+2\text{Tc} \left\langle\sigma_{x}\right\rangle\right)
\end{equation}
where
\begin{equation}
\left\langle\sigma_{x}\right\rangle=r\sin\theta\cos\phi,\quad\left\langle\sigma_{y}\right\rangle=r\sin\theta\sin\phi,
\quad\left\langle\sigma_{z}\right\rangle=r\cos\theta
\end{equation}
are the expected values of the Pauli matrices. The maximum ergotropy of a two level system is reached when:
\begin{equation}\label{max-parameters}
 r=1,\quad \sin\phi=0\quad \text{and} \quad\tan\theta=\frac{2\text{T}c}{\epsilon},
\end{equation}
and its value is
\begin{equation}\label{max_ergo}
\mathcal{W}_{\text{max}}=\Delta.
\end{equation}
In the figure \ref{fig_1} we can observe the variation of the ergotropy in spherical coordinates of the Bloch sphere for all possible values of $\rho$. As a particular case, the system can be initialized in the excited state with no coherence (nc) $\ket{0}$, which corresponds to $r=1$ and $\theta=0$   
(see equation \ref{gen_rho}). In  this case the ergotropy and the maximum ergotropy are respectively:
\begin{equation}
\mathcal{W}_{\text{nc}}=\frac{1}{2} r\left(\Delta+\epsilon\right),\quad
\mathcal{W}_{\text{nc,max}}=\frac{1}{2} \left(\Delta+\epsilon\right).
\end{equation}
Note that in general 
\begin{equation}
    \mathcal{W}\geq\mathcal{W}_{\text{nc}}
\end{equation}
which means that more work can be extracted from a system with quantum coherence than from its classical counterpart. Note that the maximum ergotropy in equation \ref{max_ergo} corresponds to the ergotropy of the excited state $\Delta =\varepsilon_2-\varepsilon_1$ (equation \ref{h_eigen}), which is to be expected we were working on the energy basis. The relationship between coherence and entropy on the energy basis is analyzed in \cite{francica2020quantum}.   

\section{Feedback control on a transport system}

\begin{figure}[ht]
\centering
\includegraphics[width=0.6\textwidth]{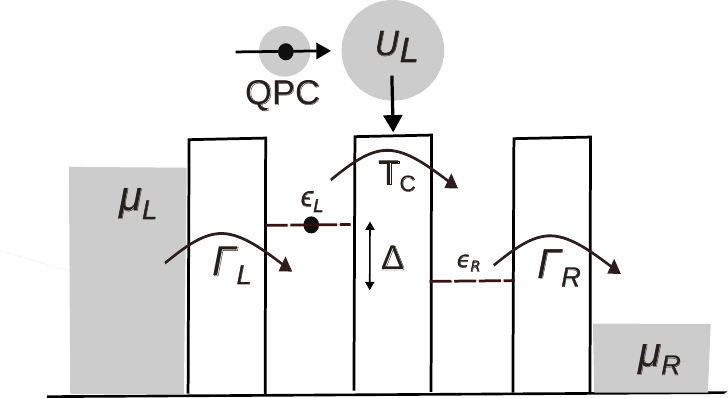}
\caption{Schematic representation of the electronic transport in a double quantum dot coupled to two electronic reservoirs in the Coulomb blockade regime, with the feedback operation and its respective parameters.}\label{fig_1.1}
\end{figure}

It seems natural to ask whether there is an effective way to reach a quantum state with maximum ergotropy for a two-level system. The most common approach is to use an external field to excite the system, but a unitary evolution does not allow to stabilize the state towards a desired one. \cite{Carrega2020, Santos2019}. Here we study the possibility of using a feedback scheme to stabilize the quantum state at a target state with optimal ergotropy. The feedback scheme used is the proposed by Wiseman and Milburn, which has been used both theoretically \cite{Wang2001} and experimentally \cite{Sayrin2011} to stabilize pure states in optical systems, and has been studied in charge transport systems. 
We start from the system proposed by Pöltl et al. \cite{Poltl2011}, in which a two-level system (a double quantum dot coupled in series) coupled to two electron reservoirs (one lead to the left L and another to the right R) at different chemical potentials is subjected to the feedback scheme, where the electron transport information is used to modify the electron transport itself (Figure \ref{fig_1.1}). 
This system has been extensively studied in works such as \cite{Gurvitz1996, Stoof1996}, and  can be described by the Markovian master equation in the occupation basis for the double quantum dot established in \cite{Gurvitz1996, Stoof1996}. This equation works under two considerations: within the high-bias limit, when the chemical potential of one reservoir is much higher than the other reservoir ($\mu_L>>\mu_R$); and with Coulomb blockade, where only a single electron is allowed in the system at any time. This allows to span the system space on three levels: the unoccupied state $\ket{0}$, and the occupied states at the left dot $\ket{L}$, and at the right dot $\ket{R}$. In the Liouville space (where calligraphic notation is used for operators), the master equation without the feedback scheme can be written in the form of a smooth evolution followed by jumps between the energy levels  
\begin{equation}
\dot{\rho}=\mathcal{L} \rho=\left(\mathcal{L}_{0}+\mathcal{J}_{L}+\mathcal{J}_{R}\right) \rho,
\end{equation}
where $\mathcal{L}$ is the total Liouvillian of the evolution, $\mathcal{L}_0$ is the free Liouvillian describing the smooth evolution of the system between jumps, and the jump operators $\mathcal{J}_{L}$, $\mathcal{J}_{R}$, describe the change of the occupation numbers of the left and right dots:     
\begin{equation}\label{Hprima}
\begin{aligned}
\mathcal{J}_{L} \rho=\Gamma_L D_{L} \rho D_{L}^{\dagger},\quad \mathcal{J}_{R} \rho=\Gamma_R D_{R} \rho D_{R}^{\dagger}\quad\text{and}\\
 \mathcal{L}_{0} \rho=-i\left(\widetilde{H} \rho-\rho \widetilde{H}^{\dagger}\right),\quad\text{with}\quad \widetilde{H}=H_{S}-i \frac{1}{2} \sum_{\alpha=L,R} \Gamma_{\alpha} D_{\alpha}^{\dagger} D_{\alpha}.
 \end{aligned}
\end{equation}
Here $\Gamma_{L},\Gamma_{R}$ are the reservoir tunnel rates and $D_L=\ketbra{L}{0}$, $D_R=\ketbra{0}{R}$, the coupling operators. $\widetilde{H}$ is an effective Hamiltonian whose left eigenvectors $\left|\phi_{j}\right\rangle$ will be the target states of the control operation. Since $\widetilde{H}$ is a non-Hermitian operator, its left eigenvectors are different from its right eigenvectors $\langle\tilde{\phi}_{j}| \neq\langle\phi_{j}|$. In its eigenbasis 
\begin{equation}
\widetilde{H}=-i \frac{\Gamma_{L}}{2}|0\rangle\langle0|+\sum_{j=1}^{N=2} \widetilde{\varepsilon}_{j}| \phi_{j}\rangle\langle\tilde{\phi}_{j}|.
\end{equation}
Because of the high-bias limit, the system has a unidirectional charge transport system from left to right:  
\begin{equation}
\mathcal{J}_L| 0\rangle\left\langle 0\left|=\Gamma_{L}\right| L\right\rangle\langle L|,\qquad 
\mathcal{J}_R| R\rangle\left\langle R\left|=\Gamma_{R}\right| 0\right\rangle\langle 0|.
\end{equation}
Then, the control operation can be applied as follows: each time an electron jumps to the left dot, it is detected by a device that triggers the control operation on the same electron by applying a unitary operation $U_L$, which is considered to be instantaneous. This type of feedback control can be implemented experimentally by placing a quantum point contact (QPC) near the left dot acting as the detector, and a pulsed gate voltage acting as the unitary control operation. The unitary control operation is such that pointing to the desired state $| \phi_{j}\rangle$ rotates the state of the electron so that 
\begin{equation}
U_L| L\rangle=|\phi_{j}\rangle,
\end{equation}
then an effective jump operator can be defined in such a way that it causes the electron to jump from the left reservoir to the target state. In the Liouville space we have 
\begin{equation}\label{jumpl}
\mathcal{J}_L^{(C)}| 0\rangle\langle 0|=
\mathcal{U}_L\mathcal{J}_L| 0\rangle\langle 0|
=\Gamma_{L} \mathcal{U}_L | L\rangle\langle L|
=\Gamma_{L}| \phi_{j}\rangle\langle \phi_{j}|.
\end{equation}
Once the state reaches the target, it relaxes by jumping the electron to the right reservoir with an effective tunneling rate $\Gamma_{R}^j=-2\text{Im}(\widetilde{\varepsilon}_j)$
\begin{equation}\label{Gamma_eff}
\mathcal{J}_R| \phi_{j}\rangle\langle \phi_{j}|=\Gamma_{R}^j| 0\rangle\langle 0|,
\end{equation}
where $\widetilde{\varepsilon}_j$ are the eigenvalues of $\widetilde{H}$ \cite{Poltl2011} (Appendix A), then the Markovian master equation with feedback control is: 
\begin{equation}\label{Mark_master_eq}
\begin{gathered}
\dot{\rho}=\mathcal{L}^{(C)} \rho=\left(\mathcal{L}_{0}+\mathcal{J}_{L}^{(C)}+\mathcal{J}_{R}\right) \rho \\
 \text{with} \qquad \mathcal{J}_{L}^{(C)}\rho=\Gamma_L D_{L} ^{(C)}\rho D_{L}^{\dagger{(C)}} \qquad \text{and} \qquad
 D_{L} ^{(C)}=U_LD_{L},
\end{gathered}
\end{equation}
which steady-state solution is 
\begin{equation}
\rho_{\mathrm{stat}}=\frac{1}{ \Gamma_{L}+\Gamma_{R}^{j}}\left(\Gamma_{R}^{j}|0\rangle\langle0|+ \Gamma_{L} |\phi_{j}\rangle\langle\phi_{j}|\right),
\end{equation}
so, the limit when $\Gamma_{R}\rightarrow 0$ implies $\Gamma_R^{j}\rightarrow 0$, and the control operation ideally allows to reach the target state
\begin{equation}\label{ss_control}
\lim _{\Gamma_{R}\rightarrow 0} \rho_{\text {stat }}=|\phi_{j}\rangle\langle\phi_{j}|.
\end{equation}
Note that in this scheme the controlled system can effectively be understood as a single resonant level model $|\phi_{j}\rangle$ connected to two electronic reservoirs, one on the left and one on the right, with transition rates $\Gamma_L$ and $\Gamma_{R}^{j}$ respectively.   

\section{Ergotropy of a controled transport qubit}

\begin{figure}[ht]
\centering
\includegraphics[width=0.5\textwidth]{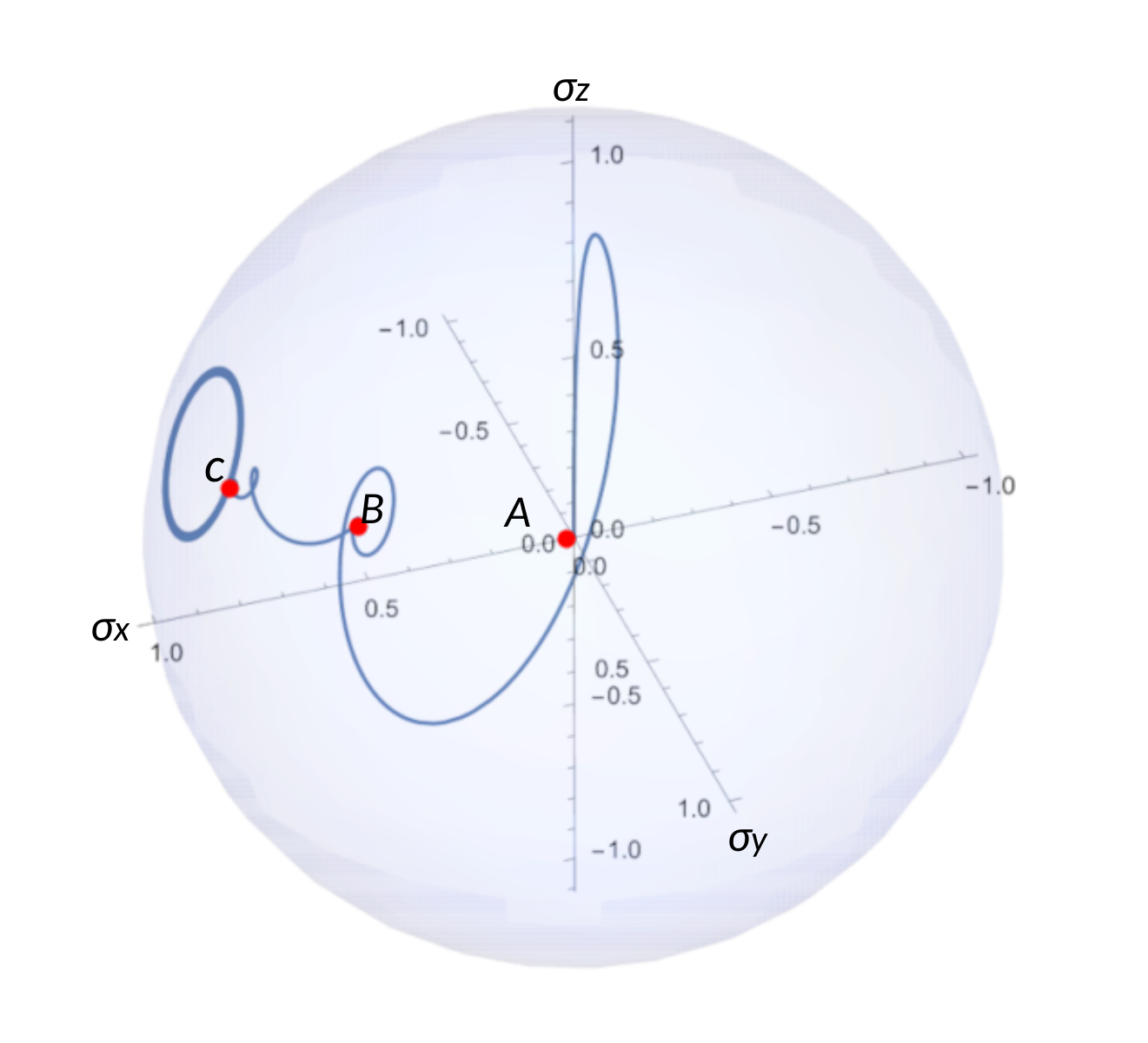}
\caption{Dynamics of the evolution of the states of the double quantum dot from the initial state A (empty state), through the steady state without control B, to the target state C, with the parameters $\epsilon=\text{Tc}=\Gamma_R$. In B the optimal control is activated and the system is decoupled from the reservoirs in C.}\label{fig_2}
\end{figure}

\begin{figure}[ht]
\centering
\begin{subfigure}{.5\textwidth}
  \centering
  \includegraphics[width=1.0\linewidth]{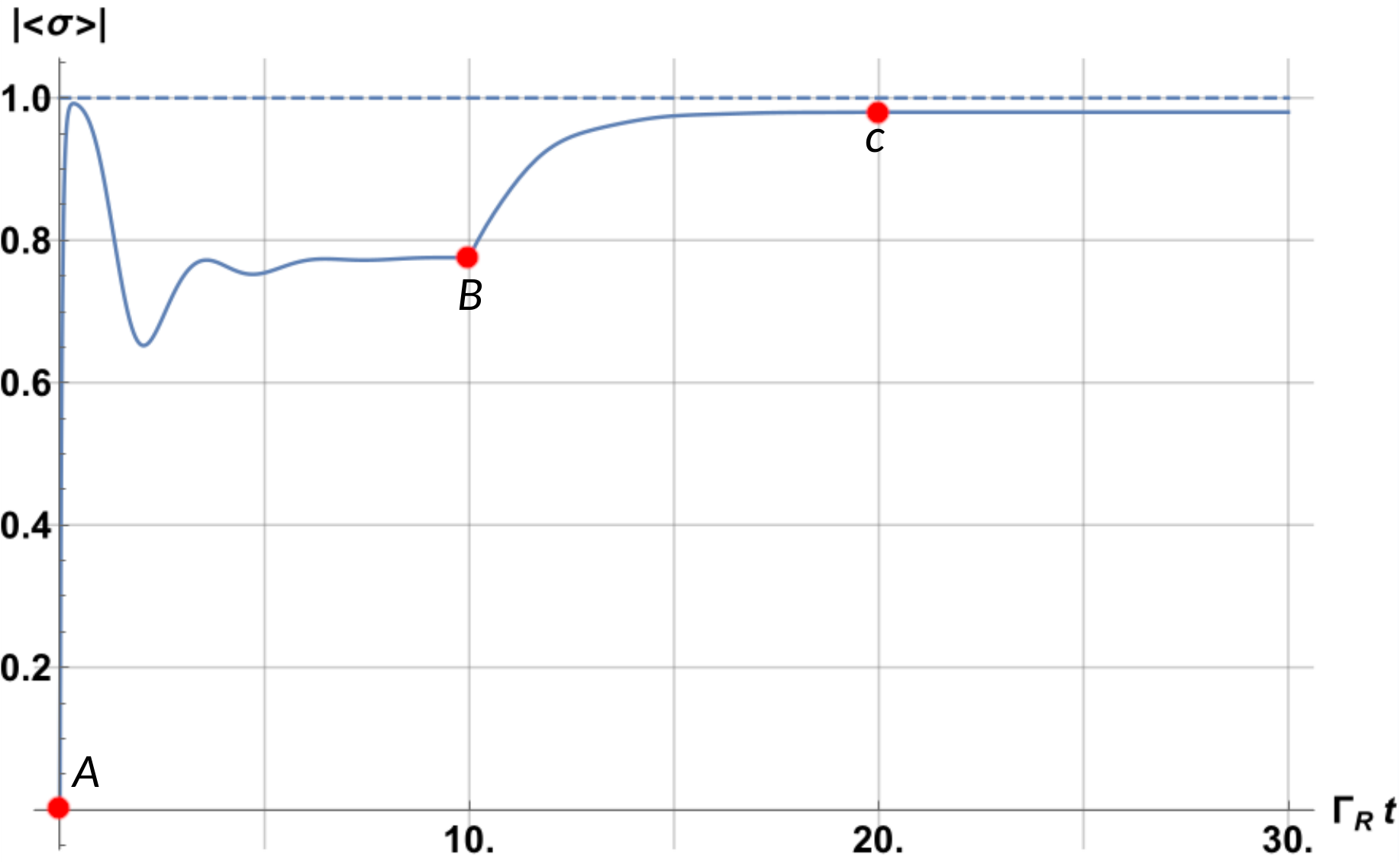}
\end{subfigure}%
\begin{subfigure}{.5\textwidth}
  \centering
  \includegraphics[width=1.0\linewidth]{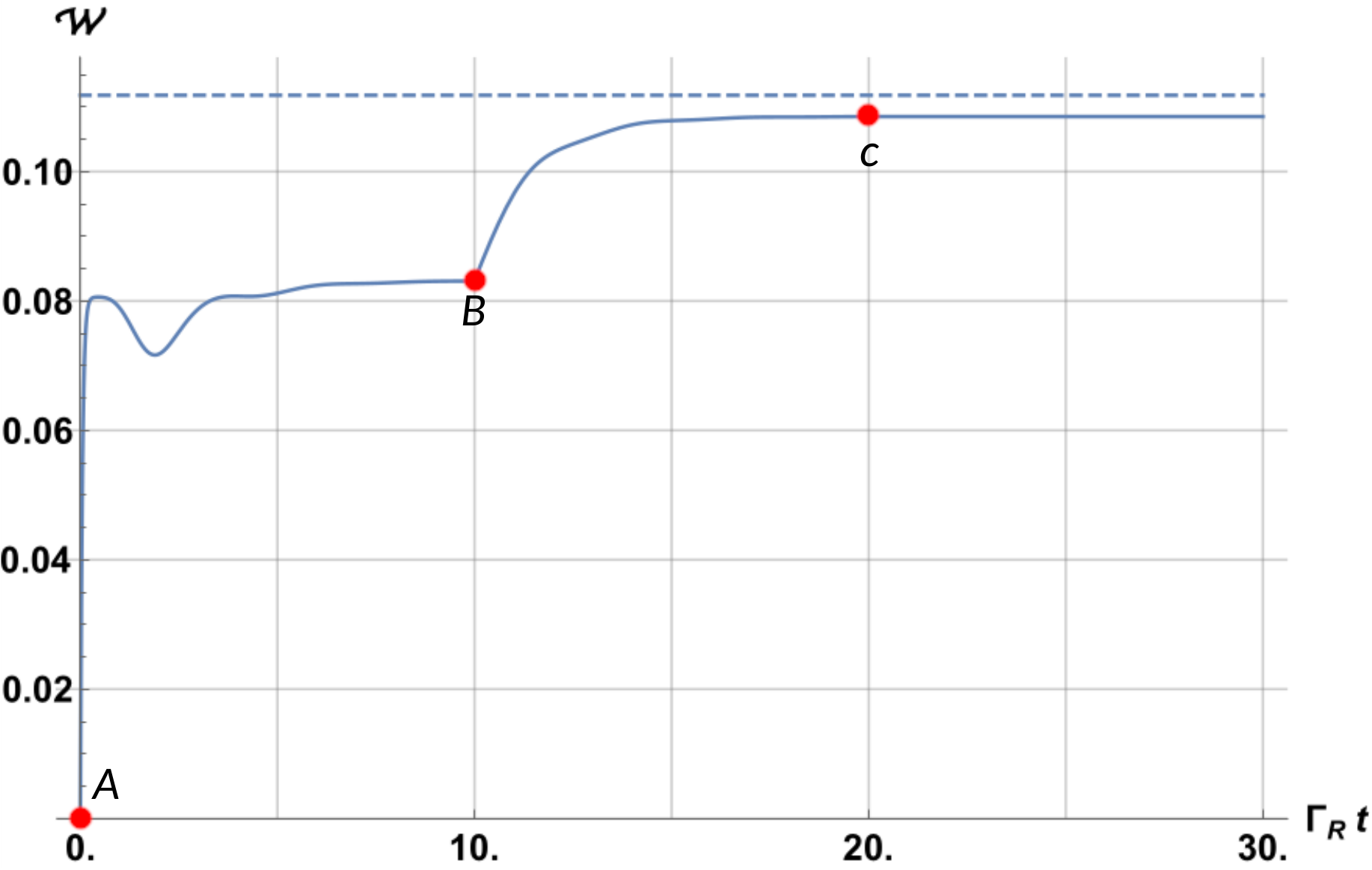}
\end{subfigure}
\caption{Variation of ergotropy and Bloch vector through the double quantum dot dynamics. Points A, B and C correspond to the times of the corresponding states shown in Figure \ref{fig_2}. The dotted lines are the maximum values for the parameters $\epsilon=\text{Tc}=\Gamma_R$.}\label{fig_3}
\end{figure}

The Hamiltonian for a double quantum dot in series is: 
\begin{equation}
 H _ { S } = \frac { \epsilon } { 2 } \sigma_z+\text{Tc}\sigma_x \quad
\text{with} \quad 
\sigma_{z}=|L\rangle\langle L| - |R\rangle\langle R|, \quad 
\sigma_{z}=|L\rangle\langle R| + |R\rangle\langle L|.
\end{equation}
To maximize the ergotropy of the system when it is coupled to two reservoirs in the scheme of the previous section, it is necessary to find the parameters of the control operation that point to the target state $|\phi_{j}\rangle$. Following Pöltl et al. \cite{Poltl2011}, it is easier to find an explicit form for $|\phi_{j}\rangle$, in the pseudospin basis 
\begin{equation}\label{pseudospin-basis}
\begin{aligned}\{\rho\}=&
\{\rho_{00}, n_{\mathrm{occ}},\left\langle\sigma_{x}\right\rangle,\left\langle\sigma_{y}\right\rangle,\left\langle\sigma_{z}\right\rangle\}\\
=&\{\rho_{00}, \rho_{L L}+\rho_{R R}, \rho_{L R}+
\rho_{R L}, \frac{1}{i}\left(\rho_{L R}-\rho_{R L}\right), \rho_{L L}-\rho_{R R}\},
\end{aligned}
\end{equation}
than in the occupation basis.
In the basis pseudospin basis the target state reads ($\Gamma_W=\sqrt{\left(4 \epsilon^2+\Gamma_R^2+16 T_C^2\right)^2-64 \Gamma_R^2 T_C^2}$)
\begin{equation}\label{sigma_coor}
\begin{aligned}
&\left\langle\sigma_{x}\right\rangle=-\frac{4 \epsilon^{2}+\Gamma_{R}^{2}-16 \text{Tc}^{2}-\Gamma_{W}}{16 \epsilon \text{Tc}}\left\langle\sigma_{z}\right\rangle, \\
&\left\langle\sigma_{y}\right\rangle=\frac{4 \epsilon^{2}+\Gamma_{R}^{2}+16 \text{Tc}^{2}-\Gamma_{W}}{8 \Gamma_{R} \text{Tc}}, \\
&\left\langle\sigma_{z}\right\rangle=\mp \frac{\sqrt{\Gamma_{R}^{2}-4 \epsilon^{2}-16 \text{Tc}^{2}+\Gamma_{W}}}{\sqrt{2} \Gamma_{R}}.
\end{aligned}
\end{equation}
The feedback control is applied using an unitary rotation operator
\begin{equation}
\hat{U}=e^{-i \theta_{c} \vec{n} \cdot \vec{\sigma}}
\end{equation}
where $\vec{\sigma}$ is the vector of Pauli matrices and $\vec{n}=\left\{n_{x}, n_{y}, n_{z}\right\}=\{\sin \theta, 0, \cos \theta\}$ is the unit vector that determines the direction around which the unitary operator rotates the state by an angle $\theta_{c}$. The control operation only affects the left jump operator, leaving the right jump operator unchanged; thus, from the equation \ref{jumpl}, the relations between the control parameters and the target state of the double quantum dot were obtained:
\begin{equation}
\begin{aligned}
&\left\langle\sigma_{x}\right\rangle=\sin (2 \theta) \sin \left(\theta_{C}\right)^{2} \\
&\left\langle\sigma_{y}\right\rangle=\sin (\theta) \sin \left(2 \theta_{C}\right) \\
&\left\langle\sigma_{z}\right\rangle=\cos (\theta)^{2}+\cos \left(2 \theta_{C}\right) \sin (\theta)^{2}
\end{aligned}
\end{equation}
with solution  
\begin{equation}
\theta =\arccos\left(\frac{\langle\sigma_x\rangle}{\sqrt{\langle\sigma_x\rangle^2+( \langle\sigma_z\rangle-1)^2}}\right),\quad\theta_C=\arccos\left(\frac{\langle\sigma_y\rangle}{\sqrt{2-2 \langle\sigma_z\rangle}}\right).
\end{equation}
As shown in \cite{Poltl2011}, half of the states of the Bloch sphere, namely those with $\left\langle\sigma_{y}\right\rangle\geq0$, can be reached by varying the parameters of the system.
One can find out under which parameters the ergotropy can be optimized by replacing the equation \ref{sigma_coor} in \ref{gen_erg1}. We found that the maximum ergotropy (equation \ref{max_ergo}) is reached just in the limit when $\Gamma_R\rightarrow 0$ (equation \ref{sigma_coor}); in this limit
\begin{equation}\label{max_bloch-val}
\langle\sigma_x\rangle=\frac{2 \text{Tc}}{\sqrt{4 \text{Tc}^2+\epsilon ^2}},\quad\langle\sigma_y\rangle=0,\quad\langle\sigma_z\rangle=\frac{\epsilon }{\sqrt{4 \text{Tc}^2+\epsilon ^2}}.
\end{equation}
It is easy to verify that by replacing the values of the equation \ref{max_bloch-val} in \ref{gen_erg} we obtain the maximum ergotropy. The limit for reaching maximum ergotropy coincides with the limit for optimizing feedback control (equation \ref{ss_control}), which means that whenever the control is applied, the state of maximum ergotropy is reached. We conclude that the control operation not only stabilizes the system states in pure states, but also leads them directly to the state of maximum ergotropy whenever that $\epsilon>0$ and $\left\langle\sigma_{x}\right\rangle>0$ as can be seen from equation \ref{gen_erg1}.
A battery charging protocol can then be set up as follows: First, the double quantum dot is placed between the two reservoirs, initializing the system in the empty state, denoted by A in Figure \ref{fig_2}, until the system reaches its steady state B without control. Then, the feedback control is activated making $\Gamma_R\rightarrow 0$, which brings the state of the system towards the target state C with maximum ergotropy. Finally, $\Gamma_R,\Gamma_L\rightarrow 0$ is imposed so that the system is decoupled from the reservoirs and remains in a state close to the maximum ergotropy state. The dynamics of the control operation can be measured by the length of the Bloch vector $|\left\langle\boldsymbol{\sigma}(t)\right\rangle|$, which is a measure of the purity of the states. 
\begin{equation}
|\left\langle\boldsymbol{\sigma}(t)\right\rangle|^{2}=\left\langle\sigma_{x}(t)\right\rangle^{2}+\left\langle\sigma_{y}(t)\right\rangle^{2}+\left\langle\sigma_{y}(t)\right\rangle^{2}.
\end{equation}
The Bloch vector is a measure of the effectiveness of the control operation because it points to the target state whenever the control is applied. Figure \ref{fig_3} shows how the Bloch vector and ergotropy change for the dynamics shown in Figure \ref{fig_2}, and we can see that the control operation is quite effective in achieving the desired state of maximum ergotropy.  

\section{Influence of a phonon environment}

\begin{figure}[ht]
\centering
\begin{subfigure}{.5\textwidth}
  \centering
  \includegraphics[width=1.0\linewidth]{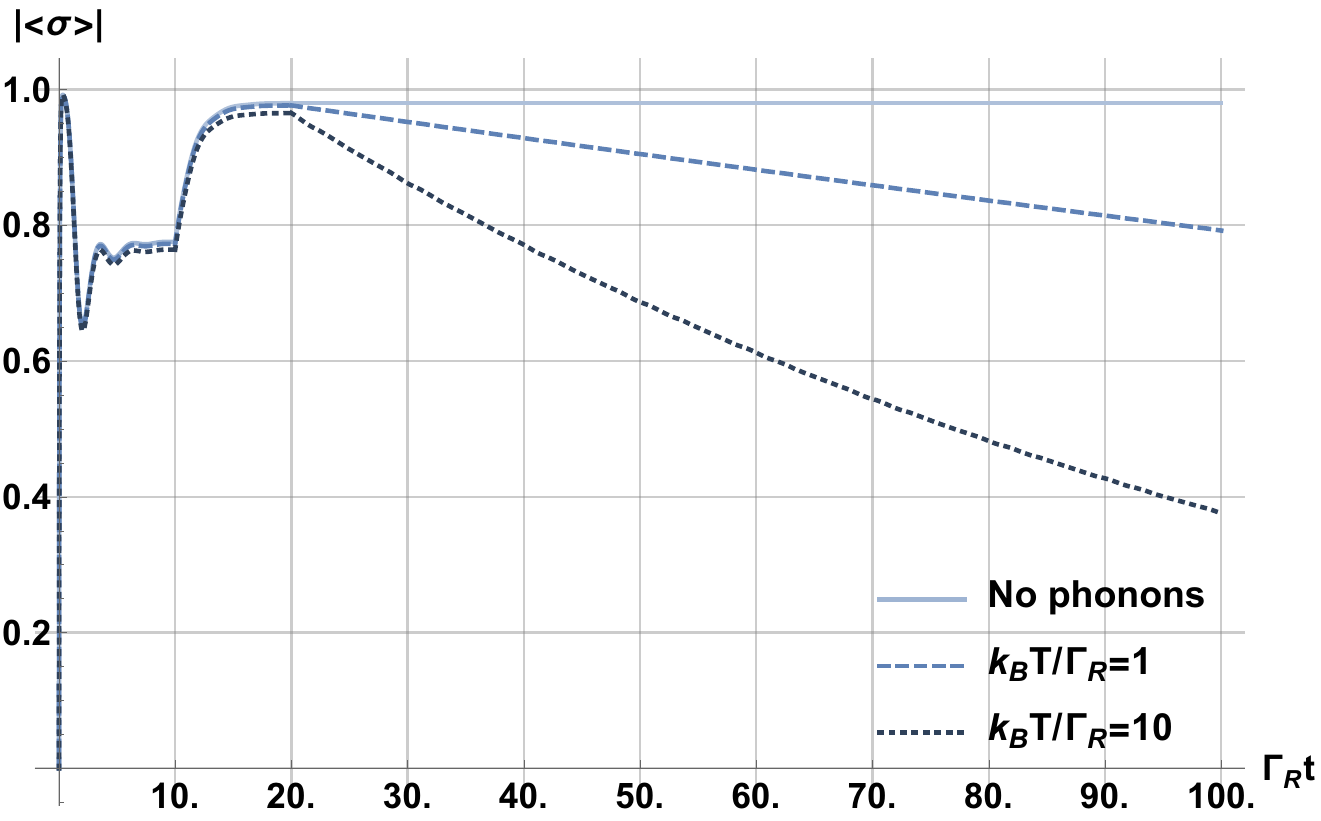}
\end{subfigure}%
\begin{subfigure}{.5\textwidth}
  \centering
  \includegraphics[width=1.0\linewidth]{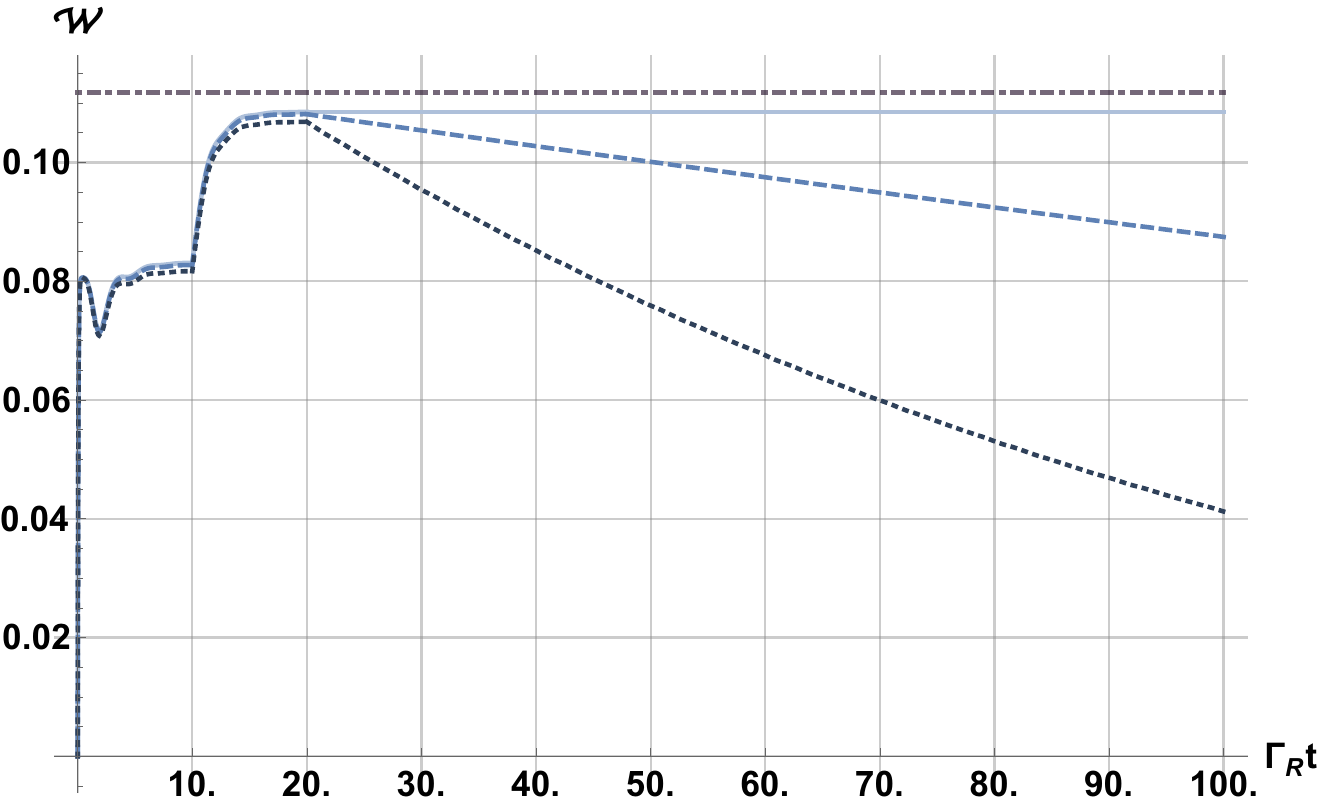}
\end{subfigure}
\caption{Variation of ergotropy and Bloch vector through the double quantum dot dynamics affected by thermal phonons at different temperatures. Parameters: $\epsilon=\text{Tc}=\Gamma_R$, $g=4\times10^{-4}$, $\omega_c/\Gamma_R=500$.}
\label{fig_4}
\end{figure}

\begin{figure}[ht]
\centering
\begin{subfigure}{.5\textwidth}
  \centering
  \includegraphics[width=1.0\linewidth]{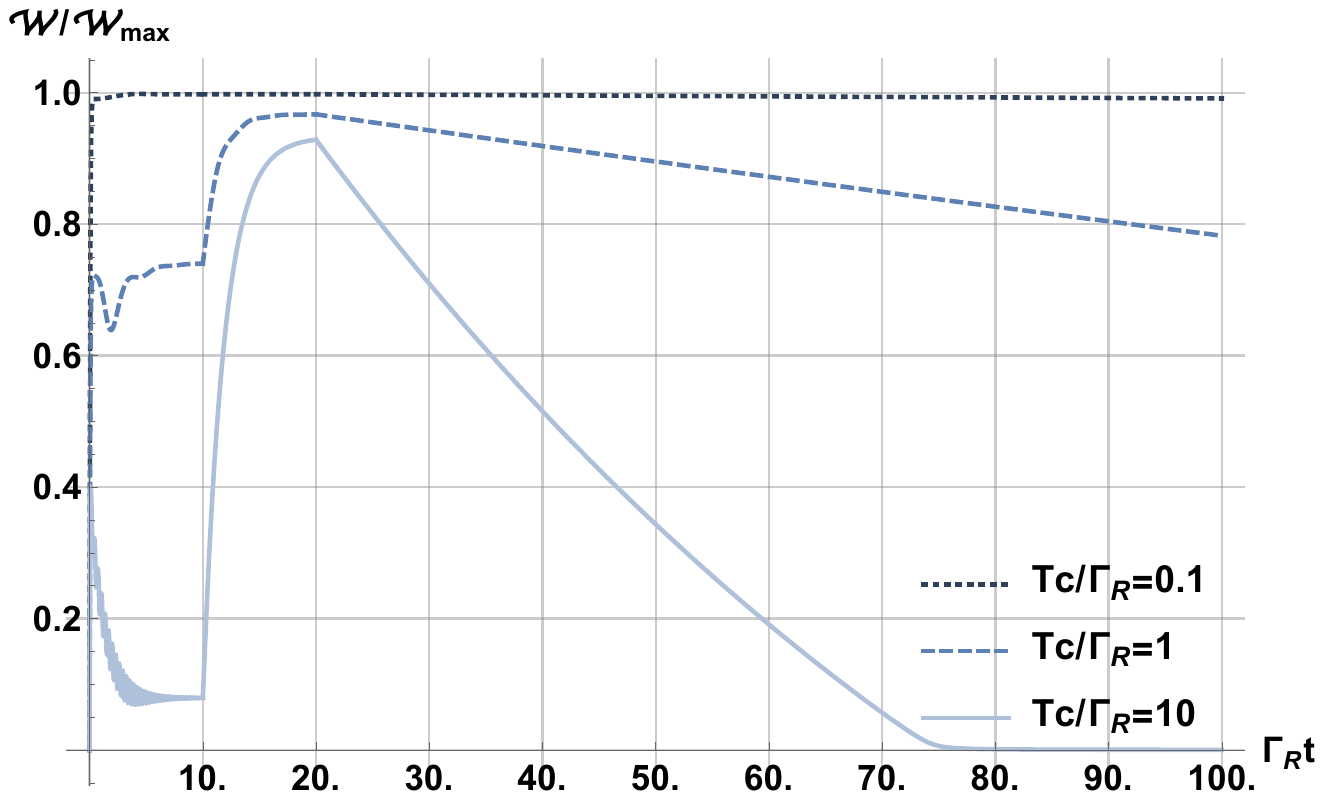}
\end{subfigure}%
\begin{subfigure}{.5\textwidth}
  \centering
  \includegraphics[width=1.0\linewidth]{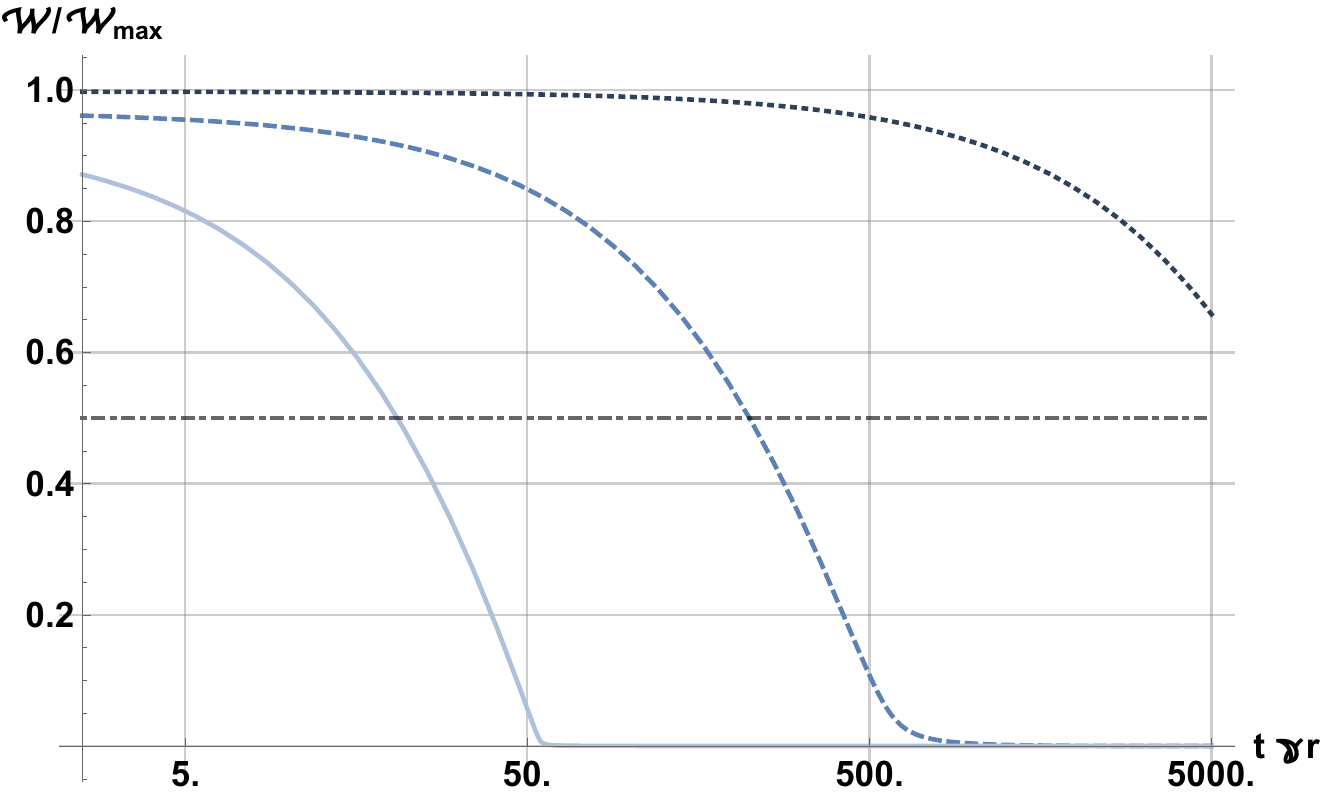}
\end{subfigure}
\caption{Variation of ergotropy through the double quantum dot dynamics for different $\text{Tc}$ parameters. Parameters: $k_B T=\Gamma_R$, $\epsilon=\Gamma_R$, $g=4\times10^{-4}$, $\omega_c/\Gamma_R=500$.}
\label{fig_6}
\end{figure}

It is important to study the interaction with a thermal phonon environment because it determines the  dissipation mechanism of the system and thus affects the charging and self-discharging processes of the battery. Self-discharge is the natural discharge process of a battery when it is not being used as a power source \cite{Santos2021}. Like any other battery, the self-discharge time is desired to be very long compared to the charging time; in fact, ideally, without interaction with the phonon environment, the discharge time of the studied system is infinite (Figure \ref{fig_3}). However, the dissipation processes are detrimental to both, the amount of charge stored and the self-discharge time, so it must be taken into account when analyzing the battery performance. The main dissipation mechanism affecting the system is the electron-phonon coupling \cite{Brandes2005};  
to include it in the master equation the phonon spectral density is assumed to have a typical ohmic parameterization  as a function of phonon frequency $\omega$:
\begin{equation}
 \rho(\omega)=g\omega e^{-\frac{\omega}{\omega_c}},
\end{equation}
where $g$ is a dimensionless coupling constant and $\omega_c$ is the Debye cutoff frequency. In the limit of weak interaction with the thermal phonon bath, $g\ll 1$, at temperature $T$, and bosonic distribution $n_B(\omega)=(e^{\beta\omega}-1)^{-1}$ with $\beta=1/(k_BT)$, the addition of $H_\text{phon}$ to the master equation yields the following dephasing rates \cite{Poltl2011} (Appendix A):
\begin{equation}
\begin{aligned}
\gamma & =g \pi T_C e^{-\Delta / \omega_c}, \\
\gamma_p & =\frac{g \pi}{\Delta^2}\left[\frac{\epsilon^2}{\beta}+2 T_C^2 \Delta e^{-\Delta / \omega_c} \operatorname{coth}\left(\frac{\beta \Delta}{2}\right)\right], \\
\gamma_b & =g \frac{\pi T_C}{\Delta^2}\left[\frac{2 \epsilon}{\beta}-\epsilon \Delta e^{-\Delta / \omega_c} \operatorname{coth}\left(\frac{\beta \Delta}{2}\right)\right].
\end{aligned}
\end{equation}
According to Poltl et al. the steady state of the system resulting from the application of the control operation is differently affected by phonons for different parameters of the system. Different branches arise as solutions of the equation $\dot{\rho}=\mathcal{L}^{(C)} \rho=0$ and they can be summarized in four: two different ones for $\epsilon$ ($\epsilon>0$ and $\epsilon<0$ respectively), and two other ones for $\sigma_x$ ($\sigma_x>0$ and $\sigma_x<0$) (see equation \ref{sigma_coor}). 
We study only the branch corresponding to $\epsilon>0$ and $\sigma_x>0$ because the others significantly reduce the ergotropy (equation \ref{gen_erg}). In Figure \ref{fig_4} we show the complete dynamic of the battery for this branch in the presence of phonons, and we see that indeed the self-discharge process is strongly affected compared to the charging process. As expected, the higher the temperature, the stronger the effect.  
\begin{figure}[ht]
\centering
\begin{subfigure}{.5\textwidth}
  \centering
  \includegraphics[width=1.0\linewidth]{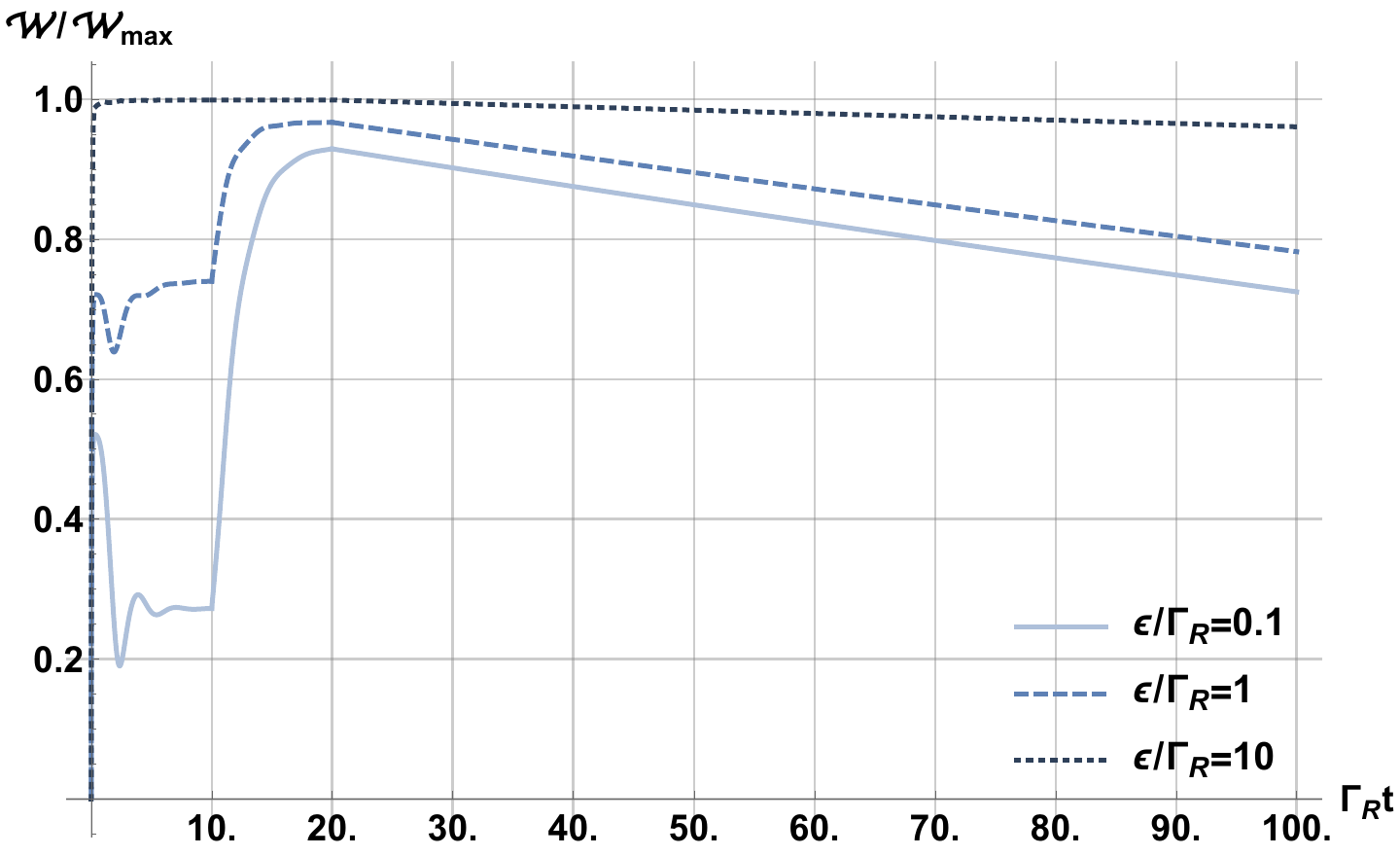}
\end{subfigure}%
\begin{subfigure}{.5\textwidth}
  \centering
  \includegraphics[width=1.0\linewidth]{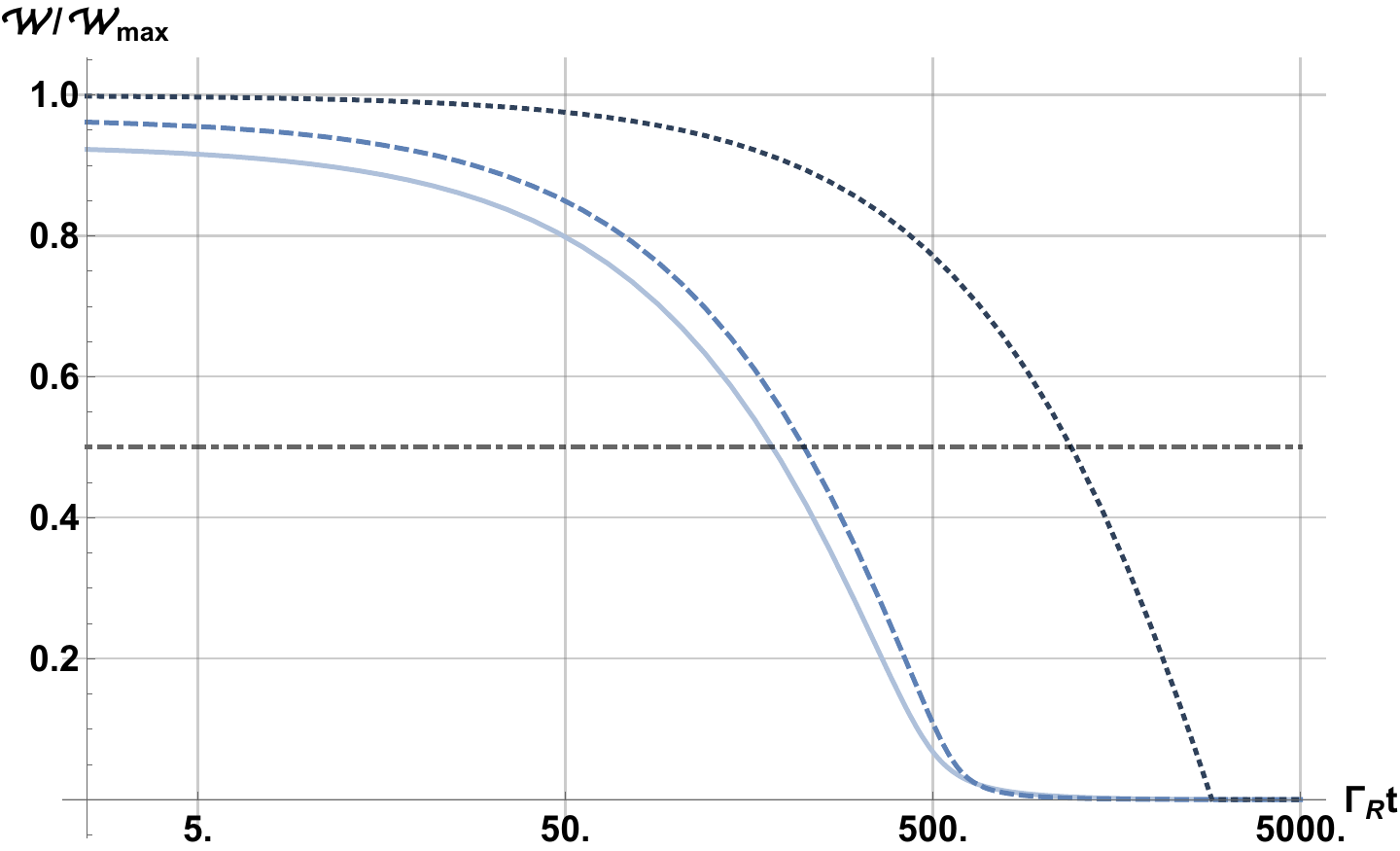}
\end{subfigure}
\caption{Variation of ergotropy through the double quantum dot dynamics for different $\epsilon$ parameters. Parameters: $k_B T=\Gamma_R$, $\text{Tc}=\Gamma_R$, $g=4\times10^{-4}$, $\omega_c/\Gamma_R=500$.}
\label{fig_5}
\end{figure}
By setting a fixed temperature, we can vary the parameters $\epsilon$ and $\text{Tc}$ to evaluate their influence on the self-discharge process, as shown in figures \ref{fig_5} and \ref{fig_6}. The left column of these figures shows the decay in logarithmic scale over time, starting from the point of maximum ergotropy. In general, we can see that at $\epsilon\gg\text{Tc}$, long discharge times are obtained, but in this limit (dotted lines) the control operation has a negligible effect on the charging, since the system reaches values close to its maximum ergotropy in very short times. This behavior results from the low coupling $\text{Tc}$ between the quantum dots, which causes the electron jumping from the left reservoir to spend a significant amount of time in the left dot before tunneling to the right dot. Thus, obtaining batteries with longer lifetimes involves trade-offs with less coherence because, according to Equation \ref{max-parameters}, in the limit when $\epsilon\gg\text{Tc}$, the polar angle would be close to zero and the coherences become small (equation \ref{gen_rho}), and therefore less stored ergotropy is obtained from coherences.

\clearpage
\section{Conclusions}
A Markovian quantum feedback protocol can be used to stabilize and achieve the maximum ergotropy in a two-level system such as the one studied in this work. The transport inducing fermionic reservoirs can be conveniently coupled or decoupled to charge the system as a quantum battery and maintain the charge over time. However, while the coupling with the bosonic environment has little effect on the charging process, it strongly affects the self-discharge of the battery. Sufficiently long discharge times are obtained when $Tc\ll\epsilon$.

\clearpage
\bibliography{biblio.bib}
\bibliographystyle{ieeetr}

\begin{appendices}
\section{}
A double quantum dot coupled in series to two electron reservoirs (one lead to the left L and another to the right R) at different chemical potentials in the high-bias limit,
in the presence of Coulomb blockade, can be described by the following Hamiltonian:
\begin{equation}\label{HamiltonianT}
 H=H_S+H_{L,R}+H_I+H_{\text {phon }},
\end{equation}
where the right-hand terms correspond to the Hamiltonian of the system, the leads in contact, the interaction between the system and the leads, and phonons respectively:
\begin{equation}
\begin{aligned}
 H_{S}=\sum_{i}^{N} \epsilon_{i} d_{i}^{\dagger} d_{i}+\sum_{i<j, j}^{N} T_{i j}\left(d_{i}^{\dagger} d_{j}+d_{j}^{\dagger} d_{i}\right),\qquad
 H_{L,R}=\sum_{k, \alpha=L, R} \epsilon_{\alpha k} c_{\alpha k}^{\dagger} c_{\alpha k},\\ \text{and} \qquad H_{I}=\sum_{k, \alpha=L, R} t_{\alpha k} c_{\alpha k}^{\dagger} d_{\alpha}+t_{\alpha k}^{*} d_{\alpha}^{\dagger} c_{\alpha k}.
\end{aligned}
\end{equation}
In general, the summation runs up to $N$ quantum dots, but for a double quantum dot system, $N=2$. Here $d_{i}^{\dagger}$ and $d_{i}$ are the creation and annihilation operators of an electron in the level $i$ with energy $\epsilon_{i}$, $T_{i j}$ is the tunneling coupling between the levels $i$ and $j$, $c_{\alpha k}^{\dagger}$ and $c_{\alpha k}$ are the creation and annihilation operators of an electron in the lead $\alpha=L,R$ in the mode $k$ with energy $\epsilon_{\alpha k}$, and $t_{\alpha k}$ is the tunneling coupling between the mode $k$ at the lead $\alpha$ with the system levels. The electron states in the lead $\alpha$ are $\left|d_{\alpha}\right\rangle=\sum_{i} \beta_{i}\left|d_{i}\right\rangle$ with amplitudes $\sum_{i}\left|\beta_{i}^{2}\right|=1$. The electron-phonon coupling
can be modeled by the interaction with a phonon bath with wave vector $Q$ and energy $\omega_Q$, through a spin-boson type Hamiltonian:
\begin{equation}
 H_{\text {phon }}=\sum_Q\left[\omega_Q a_Q^{\dagger} a_Q+\frac{1}{2} \sigma_z g_Q\left(a_{-Q}+a_Q^{\dagger}\right)\right],
\end{equation}
where $a_Q^{\dagger}$ and $a_Q$ are the phonon creation and annihilation operators, respectively, and $g_Q$ is the electron-phonon coupling strength. The Markovian master equation with feedback control, in the high-bias limit with Coulomb blockade (equation \ref{Mark_master_eq}) in the pseudospin basis (equation \ref{pseudospin-basis}) has the following operators:

\begin{equation}\label{LindO}
\mathcal{L}_{0}=\left(\begin{array}{ccccc}
-\Gamma_{L} & 0 & 0 & 0 & 0 \\
0 & -\frac{1}{2} \Gamma_{R} & 0 & 0 & \frac{1}{2} \Gamma_{R} \\
0 & -\gamma & -\frac{1}{2}\Gamma_{R}-\gamma_{p}  & \epsilon & \gamma_{b} \\
0 & 0 & -\epsilon & -\frac{1}{2}\Gamma_{R}-\gamma_{p} & 2 \text{Tc} \\
0 & \frac{1}{2} \Gamma_{R} & 0 & -2 \text{Tc} & -\frac{1}{2} \Gamma_{R}
\end{array}\right),
\end{equation}
\begin{equation}
\begin{gathered}
\mathcal{J}_{L}^{(C)}=\left(\begin{array}{ccccc}
0 & 0 & 0 & 0 & 0 \\
\Gamma_{L} & 0 & 0 & 0 & 0 \\
\Gamma_{L} \sin (2 \theta) \sin \left(\theta_{C}\right)^{2} & 0 & 0 & 0 & 0 \\
\Gamma_{L} \sin (\theta) \sin \left(2 \theta_{C}\right) & 0 & 0 & 0 & 0 \\
\Gamma_{L}\left(\cos (\theta)^{2}+\cos \left(2 \theta_{C}\right) \sin (\theta)^{2}\right) & 0 & 0 & 0 & 0
\end{array}\right)
\\
\text {and}\quad\mathcal{J}_{R}=\left(\begin{array}{ccccc}0 & \frac{1}{2} \Gamma_{R} & 0 & 0 & -\frac{1}{2} \Gamma_{R} \\ 0 & 0 & 0 & 0 & 0 \\ 0 & 0 & 0 & 0 & 0 \\ 0 & 0 & 0 & 0 & 0 \\ 0 & 0 & 0 & 0 & 0\end{array}\right).
\end{gathered}
\end{equation}

The eigenvalues of $\widetilde{H}$ used in the equation \ref{Gamma_eff} are:
\begin{equation}
 \widetilde{\varepsilon}_{\mp}=\frac{1}{4}\left(-i \Gamma_R \mp \sqrt{4 \epsilon^2+4 i \epsilon \Gamma_R-\Gamma_R^2+16 T_C^2}\right).
\end{equation}

\end{appendices}

\end{document}